\documentstyle[pra,aps,preprint,epsf]{revtex}

\begin{document}
\draft

\title{Theoretical study of the absorption spectra of the lithium dimer}
\author{H.-K. Chung, K. Kirby, and J. F. Babb}
\address{
Institute for Theoretical Atomic and Molecular Physics,\\
Harvard-Smithsonian Center for Astrophysics,\\ 
60 Garden Street, Cambridge, MA 02138}

\maketitle
%
\begin{abstract}
For the lithium dimer we calculate cross sections for absorption of
radiation from the vibrational-rotational levels of the ground
$\mbox{X}\,{}^1\Sigma_g^+$ electronic state to the vibrational levels
and continua of the excited $\mbox{A}\,{}^1\Sigma_u^+$ and
$\mbox{B}\,{}^1\Pi_u$ electronic states.  Theoretical and experimental
data are used to characterize the molecular properties taking
advantage of knowledge recently obtained from photoassociation
spectroscopy and ultra-cold atom collision studies.  The
quantum-mechanical calculations are carried out for temperatures in
the range from 1000 to 2000~K and are compared with previous
calculations and measurements.
\end{abstract}

\narrowtext
%
%
\section{INTRODUCTION}
The absorption spectra of pure alkali-metal vapors at temperatures of
the order 1000~K can be a rich source of information on molecular
potentials and properties.  Achieving a high vapor pressure of lithium
in experiments requires higher temperatures than the other
alkali-metal atoms, but there are some data from heat pipe
ovens~\cite{Vid73,VezMilPic82,TheMusDem90} and from a specialized
apparatus~\cite{ErdSanStw96}.  In addition to the atomic lines the
spectra exhibit gross molecular features attributable to transitions
between bound levels of the ground electronic state and levels of the
excited singlet states and weaker features arising from analogous
triplet transitions.

Theoretically, the envelope of the alkali-metal molecular absorption
spectra can be quantitatively reproduced using semi-classical
models~\cite{Bat52,SzuBay75,LamGalHes77,SchWoePic88,WoeSchKor85} and
ro-vibrational structure~\cite{LamGalHes77} and
continua~\cite{BucReiSpe52} can be reproduced from quantum-mechanical
models.  In this paper we calculate quantum-mechanically absorption
spectra for the $\mbox{X}\,{}^1\Sigma_g^+$--$\mbox{A}\,{}^1\Sigma_u^+$
and $\mbox{X}\,{}^1\Sigma_g^+$--$\mbox{B}\,{}^1\Pi_u$ transitions in
$\mbox{Li}_2$.  Although both semi-classical~\cite{LamGalHes77} and
quantum-mechanical~\cite{LamGalHes77,Mil94} calculations have been
performed and compared~\cite{LamGalHes77} previously for these
transitions of $\mbox{Li}_2$, recent improvements in the molecular
data prompt the present comprehensive study.

From photoassociation spectroscopy and cold collision studies
performed in the last few years as well as recent theoretical work
there have been significant critical tests of and improvements to the
molecular potentials\cite{CotDal94,AbrMcASac95,AbrMcAGer96},
particularly at long-range~\cite{MarDal95,YanDalBab97,MarAubFre97},
and transition dipole moment data~\cite{MarDal95} available for
$\mbox{Li}_2$, as well as to the value of the lifetime of the Li $2p$
state~\cite{YanDra95b,McAAbrHul96}.  This paper presents calculations
of the spectra over the full range of wavelengths where absorption in
the $\mbox{X}\,{}^1\Sigma_g^+$--$\mbox{A}\,{}^1\Sigma_u^+$ and
$\mbox{X}\,{}^1\Sigma_g^+$--$\mbox{B}\,{}^1\Pi_u$ bands is
possible. We calculate the satellite feature profiles at various
temperatures, identify and explore the influence of quasibound states
and the contributions of bound--bound versus bound--free transitions,
calculate partition functions, and calculate lifetimes for the
$\mbox{A}\,{}^1\Sigma_u^+$ and $\mbox{B}\,{}^1\Pi_u$ ro-vibrational
levels.

%
%
\section{Quantum theory of absorption cross section}

In the quantum-mechanical formulation 
an absorption cross section from a vibration-rotation state 
of the lower electronic state $(v'',J'',\Lambda'')$ 
to the vibration-rotation state of 
the upper electronic state $(v',J',\Lambda')$ is
\begin{equation}
\sigma_{v''J''\Lambda''}^{v'J'\Lambda'}(\nu)
= \frac{8\pi^3 \nu}{3hc} 
  \left|\langle\phi_{v''J''\Lambda''}\right|D(R)
  \left|\phi_{v'J'\Lambda'} \rangle\right|^2
  g(\nu-\nu_{ij}) \frac{S_{J''\Lambda''}^{J'\Lambda'}}{2J''+1}
\end{equation} 
where 
$g(\nu-\nu_{ij})$ is a line-shape function of dimension $\nu^{-1}$, 
$S_{J''\Lambda''}^{J'\Lambda'}$ is the H\"{o}nl-London factor and 
$\nu_{ij}\equiv|E_{v'J'\Lambda'}-E_{v''J''\Lambda''}|$ is 
the transition frequency ~\cite{Fri91,LefFie86}. 
In this study, $g(\nu-\nu_{ij})$ is approximated by $1/\Delta\nu$,
with $\Delta\nu$ the bin size.
For a bound-free transition, the absorption cross section from 
a bound level of the lower electronic state $(v''J''\Lambda'')$ 
to a continuum level of the upper electronic state $(\epsilon'J'\Lambda')$ 
can be written as
\begin{equation}
\sigma_{v''J''\Lambda''}^{\epsilon'J'\Lambda'}(\nu)
= \frac{8\pi^3\nu}{3hc} 
  \left|\langle\phi_{v''J''\Lambda''}\right|D(R)
  \left|\phi_{\epsilon'J'\Lambda'} \rangle\right|^2
\frac{S_{J''\Lambda''}^{J'\Lambda'}}{2J''+1}
\end{equation}
where the continuum wave function $\phi_{\epsilon' J' \Lambda'}$ is
energy normalized.  Free-bound or free-free transitions are not
considered since the temperatures studied here are not high enough for
these types of transitions to be important within the singlet
manifold.

The radial wave function  can be obtained from the
Schr\"{o}dinger equation for the relative motion of the nuclei
\begin{equation}
\label{e:schrodinger}
\frac{d^2 \phi(R)}{dR^2} + 
\left(\frac{2\mu}{\hbar^2}E - 
\frac{2\mu}{\hbar^2}V(R) - 
\frac{J(J+1)-\Lambda^2}{R^2}\right)\phi(R) = 0, 
\end{equation}
where 
$V(R)$ is the rotationless potential energy for the electronic state,
$\mu=6\,394.7$ is the reduced mass of the ${}^7\mbox{Li}$ atoms,
and $E$ is for
bound states the 
eigenvalue $E_{vJ\Lambda}$  measured with respect to 
the dissociation limit associated
with the wave function $\phi (R)=\phi_{vJ\Lambda} (R)$.
Similarly, for continuum states $E$ is the relative kinetic energy of
the colliding atoms $E_{\epsilon J \Lambda}$ associated with 
the energy-normalized wave function $\phi (R)=\phi_{\epsilon J\Lambda} (R)$.

The total absorption cross section at frequency $\nu$ can be
obtained by averaging over initial vibration-rotation levels
$(v'',J'',\Lambda'')$ with a relevant weighting factor and 
summing over all possible bound-bound and bound-free transitions with  
frequencies between $\nu$ and $\nu+\Delta\nu$~\cite{LamGalHes77} yielding
\begin{eqnarray}
\label{e:lam}
 \sigma(\nu)  = 
& Z_l^{-1}
& \left[
  \sum_{J'J''v'v''}
 \sigma_{v''J''\Lambda''}^{v'J'\Lambda'}
 \omega_{J''}
 (2J''+1)\exp[-(D_e+E_{v''J''\Lambda''})/kT] \right. \nonumber \\ 
&& \left. +\sum_{J'J''v''} \int d\epsilon' 
\sigma_{v''J''\Lambda''}^{\epsilon'J'\Lambda'}
 \omega_{J''}
 (2J''+1)\exp[-(D_e+E_{v''J''\Lambda''})/kT]
\right] ,
\end{eqnarray}
where 
$\omega_{J''}$ is a statistical factor 
due to nuclear spin with the values 
$[I/ (2I+1)]=\case{3}{8}$ for even $J$ and 
$[(I+1)/ (2I+1)]=\case{5}{8}$ for odd  $J$, 
for ${}^7\mbox{Li}_2$ with $I=\case{3}{2}$.
With the zero of energy taken to 
be the potential minimum, 
the partition function $Z_l$ of the lower state 
with  dissociation energy $D_e$ is
\begin{equation}
\label{e:partitionQM}
Z_l  =  \sum_{J''v''} \omega_{J''}(2J''+1)
     \exp[-(D_e+E_{v''J''\Lambda''})/kT] ,
\end{equation}
assuming  thermodynamic equilibrium.
The resulting cross sections
can be used to model the dimer absorption
spectra in the quasistatic limit.

%
%
\section{Molecular Data}

\begin{figure}
\centering
\epsfxsize=1.\textwidth \epsfbox{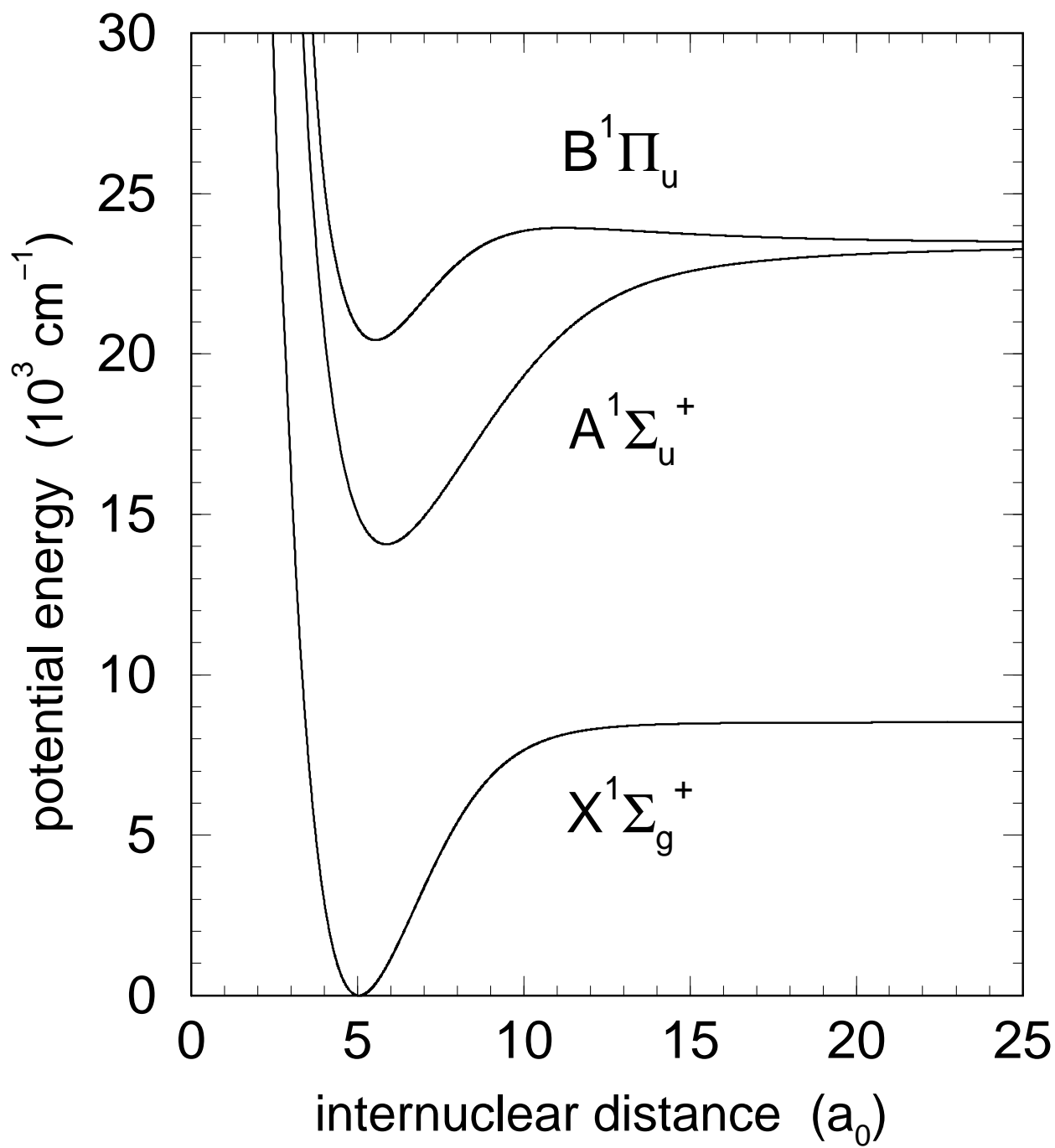}
\caption{Adopted potentials for the $\mbox{X}\,{}^1\Sigma_g^+$,
$\mbox{A}\,{}^1\Sigma_u^+$, and $\mbox{B}\,{}^1\Pi_u$ electronic
states.\label{potentials}}
\end{figure}

The adopted molecular potentials of the $\mbox{X}\,{}^1\Sigma_g^+$,
$\mbox{A}\,{}^1\Sigma_u^+$ and $\mbox{B}\,{}^1\Pi_u$ states are shown
in Fig.~\ref{potentials}.  The ground X$^1\Sigma_g^+$ state potential
was constructed using the approach of Ref.~\cite{CotDal98}.  We
adopted the recommended~\cite{ZemStw93} potential obtained by Barakat
{\em et al.}~\cite{BarBacCar86}, who applied the Rydberg-Klein-Rees
(RKR) method to measured energies.  The data were connected smoothly
to the long-range form
\begin{equation}
V(R) = -\frac{C_6}{R^6} -\frac{C_8}{R^8} -\frac{C_{10}}{R^{10}} 
    + V_{\rm exc} (R) ,
\end{equation} 
where $V_{\rm exc}(R)$ is the exchange energy~\cite{CotDal98} and the
coefficients $C_6$, $C_8$, and $C_{10}$ have been calculated in
Refs.~\cite{MarDal95,BusAubFre85,YanBabDal96,RerBusFre97} and we use
atomic units in this section. We adopted the coefficients
$C_6=1\,393.39$, $C_8=83\,425.8$ and $C_{10}=7.372\times 10^6$ from
Ref.~\cite{YanBabDal96}.  The two regions were connected at
$R=23.885\,a_0$ yielding a value of 8516.95~$\mbox{cm}^{-1}$ for the
dissociation energy of the X${}^1\Sigma_g^+$ state of $^7$Li$_2$, in
satisfactory agreement with the accepted value of
8516.61~$\mbox{cm}^{-1}$~\cite{ZemStw93}.  The form $a\exp(-bR)$ was
used to extrapolate the potential at short-range where the constants
were determined to smoothly connect to the innermost RKR points.  The
resulting $\mbox{X}\,{}^1\Sigma_g^+$ potential yields an $s$-wave
scattering length of $33.6~a_0$, in excellent agreement with the
accepted~\cite{WeiBagZil99} value of $33\pm 2$ and a sensitive test of
the assembled data.

For the excited A$^1\Sigma_u^+$ state the RKR potential of
Ref.~\cite{MarAubFre97} was adopted and smoothly connected at about
$R=140~a_0$ to the long-range form
\begin{equation}
\label{e:Along-range}
V(R) = \frac{C_3}{R^3} -\frac{C_6}{R^6} -\frac{C_8}{R^8},
\end{equation} 
with coefficients $C_3=-11.000\,226$ and $C_6=2\,075.05$ from
Ref.~\cite{YanBabDal96} and $C_8=2.705\times 10^5$ from
Ref.~\cite{MarDal95}.  For $R<3.78~a_0$ the data were connected to the
short range form $a\exp(-bR)$.  The dissociation energy for the
A${}^1\Sigma_u^+$ state is determined to be
$9\,352.194$~$\mbox{cm}^{-1}$ in our calculation, in good agreement
with the experimental value~\cite{KusHes77} of
$9\,352.5(6)$~$\mbox{cm}^{-1}$ given in Ref.~\cite{SchMinMul85}.

The potential for the B$^1\Pi_u$ state has a hump with maximum at
$R\approx 11~a_0$ and various determinations of the hump location and
height have been summarized in Refs.~\cite{OlsKon77} and
\cite{SchMinMul85}.  We adopted the IPA (Inverted Perturbation
Approach) potentials from Hessel and Vidal~\cite{HesVid79} for
$R<9.35~a_0$ and the {\it ab initio} potential of Schmidt-Mink {\em et
al.\/}~\cite{SchMinMul85} for $10.5<R<30~a_0$ with one additional
point at $R=11.2~a_0$ fixing the barrier maximum energy at
$512~\mbox{cm}^{-1}$ above dissociation.  At $R=30~a_0$ the data were
connected to the long-range form of Eq.~(\ref{e:Along-range}) by
shifting down the data by $0.3~\mbox{cm}^{-1}$ from $10.5<R<30$. The
values of the coefficients used in Eq.~(\ref{e:Along-range}) were
$C_3=5.500\,113$ and $C_6=1406.08$ from Ref.~\cite{YanBabDal96} and
$C_8=4.756\times 10^4$ from Ref.~\cite{MarDal95}.  The potential
energy data for $R<9.35~a_0$ were fixed using the
$\mbox{B}\,{}^1\Pi_u$ state dissociation energy of
$2\,984.8$~$\mbox{cm}^{-1}$, which we determined using the
experimental value for $T_e$ of
$20\,436.32$~$\mbox{cm}^{-1}$~\cite{HesVid79}, the atomic asymptotic
energy of $14\,904.0$~$\mbox{cm}^{-1}$, and the
$\mbox{X}\,{}^1\Sigma_g^+$ dissociation energy of
$8\,516.61$~$\mbox{cm}^{-1}$~\cite{ZemStw93}.  Finally, the data in
the range $9.35<R<10.5$ were smoothly connected using cubic splines.
For $R<4.254\,61~a_0$ the data were extrapolated using the short range
form $a\exp(-bR)$.

Transition dipole moments for the $\mbox{X}\,{}^1\Sigma_g^+$--$\mbox{A}\,{}^1\Sigma_u^+$ and $\mbox{X}\,{}^1\Sigma_g^+$--$\mbox{B}\,{}^1\Pi_u$
transitions
are available from {\em ab initio\/}
calculations~\cite{RatFisKon87,KonRosHoc83,SchMinMul85} and 
for X--A transitions from 
measured lifetimes of A state levels~\cite{BauKorPre84}.  
For the electronic transition
dipole moments $D(R)$, we adopted {\it ab initio\/} calculations of
Ratcliff, Fish, and Konowalow~\cite{RatFisKon87} connected at
$R=35~a_0$ to the long-range asymptotic form
\begin{equation}
D_\infty (R)= D_0 + \frac{b}{R^3}.
\end{equation}
The value of the coefficient $D_0$ was $3.3175$ for both X--A and X--B
transitions and the coefficient $b$ was $283.07$ or
$-141.53$~\cite{MarDal95} for, respectively, X--A or X--B transitions.
For both transitions, we multiplied $D_\infty(R)$ calculated using the
above coefficients by a constant such that the value $D_\infty(35)$
was identical to the corresponding {\it ab initio\/} value from
Ref.~\cite{RatFisKon87} to provide a smooth connection between short
and long-range forms for $D(R)$.  The X--A dipole moment function that
we adopted is consistent with that derived by Baumgartner {\em et
al.\/}~\cite{BauKorPre84} from experimental measurements. There is no
experimentally-derived dipole moment function for the X--B transition.

%
%
\section{Results}
Bound and continuum wave functions were calculated using the Numerov
method to integrate Eq.~(\ref{e:schrodinger}).  For the X and A
states, eigenvalues were generally in good agreement with the
Rydberg-Klein-Rees values used as input to the potentials constructed.
[There is an apparent misprint for the energy of the $v''=9$ level of
the X$^1\Sigma_g^+$ state in Ref.~\cite{BarBacCar86}.  We used
$3\,098.641\,2$ $\mbox{cm}^{-1}$, consistent with
Ref.~\cite{KusHes77}.]  The constructed $\mbox{B}\,{}^1\Pi_u$ state
potential can reproduce the rotationless IPA energies tabulated by
Hessel and Vidal typically to about $0.1$~$\mbox{cm}^{-1}$, with the
greatest discrepancy 0.15~$\mbox{cm}^{-1}$ for the $v'=13$ value.
Calculated frequencies of
$\mbox{B}\,{}^1\Pi_u$--$\mbox{X}\,{}^1\Sigma_g^+$ transitions were
also compared with calculations of Verma, Koch, and
Stwalley~\cite{VerKocStw83} and the agreement was good, within
0.1~$\mbox{cm}^{-1}$.  Four quasi-bound levels of the
$\mbox{B}\,{}^1\Pi_u$ state were found.  For the rotationless
potential their calculated eigenvalues are 143.91, 276.7, 391.74 and
483.88 $\mbox{cm}^{-1}$ above the dissociation limit for $v'=14$ to
$v'=17$.

\begin{figure}
\centering
\epsfxsize=1.\textwidth \epsfbox{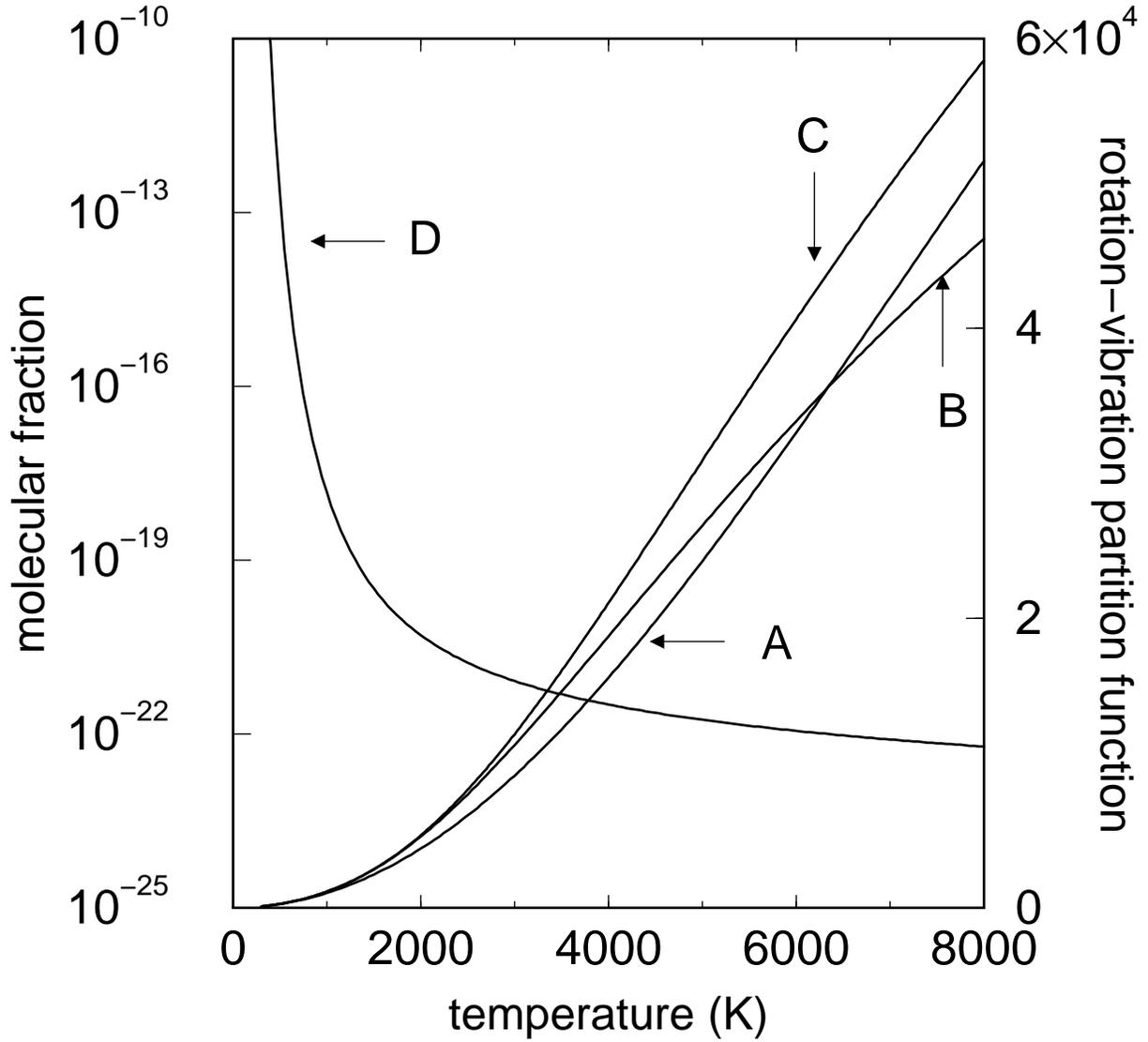}
\caption{Comparison at various temperatures of the partition functions
$\tilde{Z}_l$, calculated using Eq.~(\protect\ref{e:partitionLam}) and
experimentally determined spectroscopic constants, curve A, and $Z_l$,
from Eq.~(\protect\ref{e:partitionQM}) and numerically determined
eigenvalues, curves B and C.  Inclusion of quasibound states in the
calculation of $Z_l$ results in curve C as discussed in the text.
Curve D represents the molecular fraction $[N_{{\rm Li}_2}]/[N_{\rm
Li}]^2$, Eq.~(\ref{e:mol-frac}), as a function of temperature.
\label{partition}}
\end{figure}

The calculated term energies of vibration-rotation states in the
$\mbox{X}\,{}^1\Sigma_g^+$ state were used to compute partition
functions using Eq.~(\ref{e:partitionQM}).  The maximum vibrational
and rotational quantum numbers in our calculations are 41 and 123,
respectively, for the $\mbox{X}\,{}^1\Sigma_g^+$ state.  In the
harmonic approximation for ro-vibrational energies, the partition
function can be calculated using the simple expression
\begin{eqnarray}
\label{e:partitionLam}
 \tilde{Z_l} \equiv  Z_RZ_v  &\approx
&\sum_{J''=0}^{123} \omega_{J''}(2J''+1)\exp[-hcB_eJ''(J''+1)/kT]\nonumber \\
&&\times \sum_{v''=0}^{41} \exp[-h\nu_e(v''+1/2)/kT] ,
\end{eqnarray}
which assumes that the term energy can be described by the first terms
of the power series with respect to vibrational and rotational quantum
numbers.  Using constants $\nu_e/c$ = 351.3904 $\mbox{cm}^{-1}$ and
$B_e = 0.672\,566$ $\mbox{cm}^{-1}$~\cite{HesVid79} the partition
function from Eq.~(\ref{e:partitionLam}) was calculated and it is
compared with the partition function calculated from
Eq.~(\ref{e:partitionQM}) for the $\mbox{X}\,{}^1\Sigma_g^+$ state as
a function of temperature in Fig.~\ref{partition}.  The anharmonicity
of the potential for higher vibrational levels accounts for the
differences between the two results with increasing temperature.  For
$J>2$ the $\mbox{X}\,{}^1\Sigma_g^+$ state supports quasibound
vibrational levels.  The expression Eq.~(\ref{e:partitionQM}) for
$Z_l$ does not specify whether quasibound states are to be included or
not in the summations. We evaluated $Z_l$ with and without the
quasibound levels to ascertain their importance and the results are
shown in Fig.~\ref{partition}. The effect of the additional levels
becomes increasingly significant with higher temperature.  For the
present study covering temperatures between 1000 and 2000~K there is
not a significant distinction between $Z_l$, $\tilde{Z_l}$, and the
result with the inclusion of the quasibound states.

The molecular fraction can be calculated
using the expression 
\begin{equation}
\label{e:mol-frac}
[N_{{\rm Li}_2}]/[N_{\rm Li}]^2 
       =  (Q_{{\rm Li}_2}/Q_{\rm Li}^2) \exp(D_e/kT) ,
\end{equation}
where the atomic partition function $Q_{\rm Li}$ is $2(2\pi m_{\rm
Li}kT/h^2)^{3/2}$, with the electronic partition function for the atom
well-approximated by the spin degeneracy of 2 for the temperatures
studied in the present work, and the molecular partition function
$Q_{{\rm Li}_2}$ is $(2\pi m_{{\rm Li}_2}kT/h^2)^{3/2}Z_l$, with the
electronic partition function for the $\mbox{X}\,{}^1\Sigma_g^+$ state
taking the value 1.  The molecular fraction Eq.~(\ref{e:mol-frac}) is
plotted in Fig.~\ref{partition}.  The absorption coefficient $k(\nu)$
can be obtained if the atomic density is known from
\begin{equation}
\label{e:qm_coeff}
 k(\nu) = [N_{{\rm Li}_2}] \sigma(\nu) .
\end{equation}

%
%
\subsection{Lifetimes}

Lifetimes of the various ro-vibrational levels of the
$\mbox{A}\,{}^1\Sigma_u^+$ state were measured ~\cite{BauKorPre84} and
calculated, see for example, Refs.~\cite{Wat77,SanKurEla84,CotDal98}.
We calculated spontaneous emission transition probabilities and
lifetimes of rotational-vibrational levels of the A$^1\Sigma_u^+$
state in order to test the adopted transition dipole moment.
The spontaneous
emission rate from a bound state ($v'J'\Lambda'$) to a bound state
($v''J'\Lambda''$) is
\begin{equation}
 A(v'J'\Lambda';v''J'\Lambda'')
= \frac{64\pi^4 \nu^3}{3hc^3} g
  \left|\langle\phi_{v''J'\Lambda''}\right|D(R)
  \left|\phi_{v'J'\Lambda'} \rangle\right|^2,
\end{equation} 
where the electronic state degeneracy 
is 
\begin{equation}
g = \frac{(2-\delta_{0,\Lambda' + \Lambda''})}{2-\delta_{0,\Lambda'}} 
\end{equation}
and we have neglected change in the rotational quantum number. 
The total spontaneous emission rate from the upper level 
($v'J'\Lambda'$) can be obtained by summing over all possible
transitions to bound and continuum states
\begin{equation}
A(v'J'\Lambda') = \sum_{v''\Lambda''}A(v'J'\Lambda';v''J'\Lambda'') 
   + \sum_{\Lambda''} \int d\epsilon'' 
  A(v'J'\Lambda';\epsilon''J'\Lambda'') ,
\end{equation}
where $A(v'J'\Lambda';\epsilon''J'\Lambda'')$ is 
the spontaneous emission probability to
a continuum energy $\epsilon''$ with partial wave $J'$.  
The lifetime is
\begin{equation}
\tau = 1/A(v'J'\Lambda').
\end{equation}

\begin{figure}
\centering
\epsfxsize=1.\textwidth \epsfbox{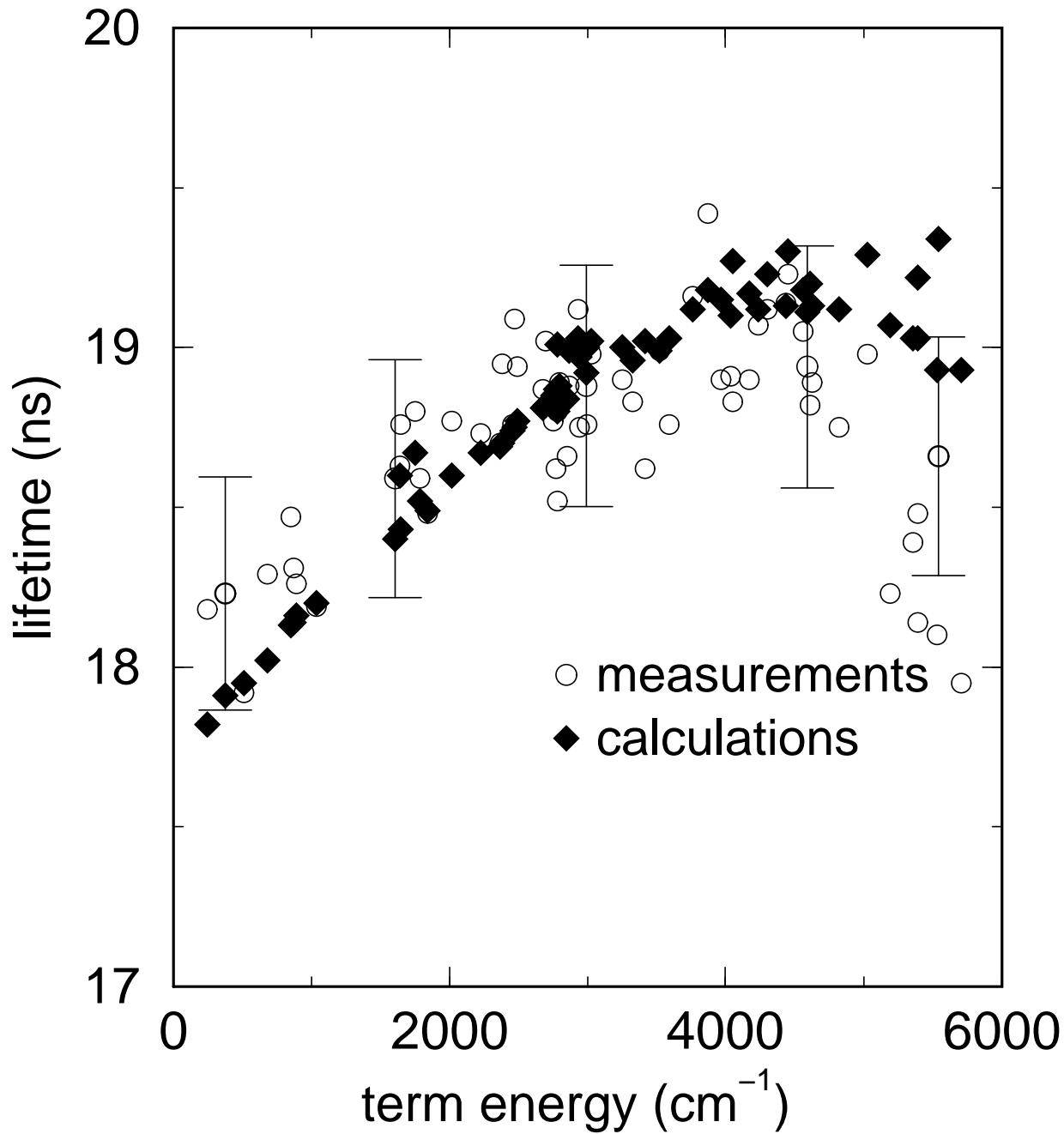}
\caption{Comparison of our calculated lifetimes (diamonds) and
measured lifetimes (circles) from Ref.~\protect\cite{BauKorPre84}.
The error bars indicate the quoted experimental uncertainty of $\pm 2$
percent.  The levels given are those that were measured, ordered by
increasing energy, as listed in Table~1 of
Ref.~\protect\cite{BauKorPre84}.
\label{flifetimes}}
\end{figure}

Lifetimes of levels $v'J'$ of the $\mbox{A}\,{}^1\Sigma_u^+$ state
were measured by Baumgartner {\em et al.\/}~\cite{BauKorPre84}.  The
$\mbox{A}\,{}^1\Sigma_u^+$ state is affected by indirect
predissociation via the ${\mbox{a}\,{}^3\Sigma_u^+}$ and
${1\,{}^3\Pi_u^+}$ states~\cite{UzeDal80,SchMinMey85}.  The measured
lifetimes $\tau_m$ of vibration-rotation levels thought to be
unaffected by indirect predissociation taken from
Ref.~\cite{BauKorPre84} and corresponding calculated lifetimes
$\tau_c$ are presented in Fig.~\ref{flifetimes} along with calculated
term energies expressed relative to the potential minimum of the A
state.  The energies are plotted in the order of the values listed in
Table~1 of Ref.~\cite{BauKorPre84} and correspond to a range of values
of $(v',J')$ from (0,15) to (24,25).  The agreement between $\tau_c$
and $\tau_m$ is good within the experimental precision of $\pm 2$
percent~\cite{BauKorPre84}.  The quasi-bound levels and continua of
the $\mbox{X}\,{}^1\Sigma_g^+$ state inside or above the high
centrifugal potential barriers are found to be important in
calculating lifetimes of high $J$ levels.  For instance, by including
transitions to three quasi-bound levels and the continuum states the
calculated lifetime of the $\mbox{A}\,{}^1\Sigma_u^+$ $(v'=20, J'=50)$
level changes from 22.7~ns to 19.3~ns which is close to the measured
lifetime of $18.66 \pm 0.37$~ns \cite{BauKorPre84}.  The calculated
and measured lifetimes agree well up to about
$4\,500$~$\mbox{cm}^{-1}$ in agreement with the findings of
Ref.~\cite{SchMinMul85}.  From approximately $5\,000$~$\mbox{cm}^{-1}$
and higher the experimental lifetimes slightly decrease relative to
theory by about 0.8~ns or five percent, as demonstrated in
Fig.~\ref{flifetimes}.  We investigated whether the $2^1\Sigma_g^+$
state might supply an additional spontaneous decay channel, but the
theoretical value $T_e=20\,128~\mbox{cm}^{-1}$~\cite{SchMinMul85} for
this state appears to place its minimum at around
$6\,000$~$\mbox{cm}^{-1}$ relative to the minimum of the
$\mbox{A}\,{}^1\Sigma_u^+$ state, apparently ruling this mechanism
out.  The reason for the significant downturn of experimental
lifetimes for higher term energies is currently not understood;
however, the overall excellent agreement between our calculated
lifetimes and the measurements gives us confidence in our molecular
data and calculational procedures.  We compare selected examples from
the present results with calculations by Watson~\cite{Wat77} and by
Sangfelt {\em et~al.\/}~\cite{SanKurEla84} in Table~\ref{t:Alife}.
The calculations of Sangfelt {\em et al.\/} are larger than ours,
probably because theoretical transition energies were used. A simple
rescaling using experimental energies, as pointed out by those
authors, yields lifetimes in good agreement with the present work.
The dipole moment function calculated by Watson~\cite{Wat77} is in
good agreement with that adopted in the present study and cannot
account for the shorter lifetimes obtained in that study.  We present
a more extensive tabulation of $\mbox{A}\,{}^1\Sigma_u^+$ lifetimes in
Table~\ref{t:Alife2} covering the same values tabulated in Table~VII
of Ref.~\cite{SanKurEla84}.

Lifetimes for levels of the $\mbox{B}\,{}^1\Pi_u$ state have been
calculated by Uzer, Watson, and Dalgarno~\cite{UzeWatDal78}, Uzer and
Dalgarno~\cite{UzeDal80}, and Sangfelt {\em et
al.\/}~\cite{SanKurEla84} and there appear to be no experimental data.
In Table~\ref{t:tlifetimes} we compare our calculated lifetimes of
selected $\mbox{B}\,{}^1\Pi_u$ levels with available calculations for
higher values of $J'$.  The present results lie between the
calculations of Uzer {\em et al.\/}~\cite{UzeWatDal78} and those of
Sangfelt {\em et al.\/}~\cite{SanKurEla84}.  The lifetimes calculated
by Uzer {\em et al.\/} are longer than ours because their transition
dipole moment function, calculated using a model potential method, is
smaller than the {\em ab initio\/} dipole moment of Ratcliff {\em
et~al.\/}~\cite{RatFisKon87} adopted in the present study.  The dipole
moment function calculated by Sangfelt {\em et al.\/}, on the other
hand, is in good agreement with that adopted in the present work.  As
those authors pointed out, and as our results illustrate, the
utilization in their calculation of calculated excitation energies
which were larger than experimental energies yielded lifetimes that
were too short.  In Table~\ref{t:tlifetimes2} we present a more
extensive tabulation of $\mbox{B}\,{}^1\Pi_u$ lifetimes covering the
same values tabulated in Table~VIII of Ref.~\cite{SanKurEla84}.

%
%
\subsection{Absorption cross sections}

\begin{figure}
\centering
\epsfxsize=.9\textwidth\epsfbox{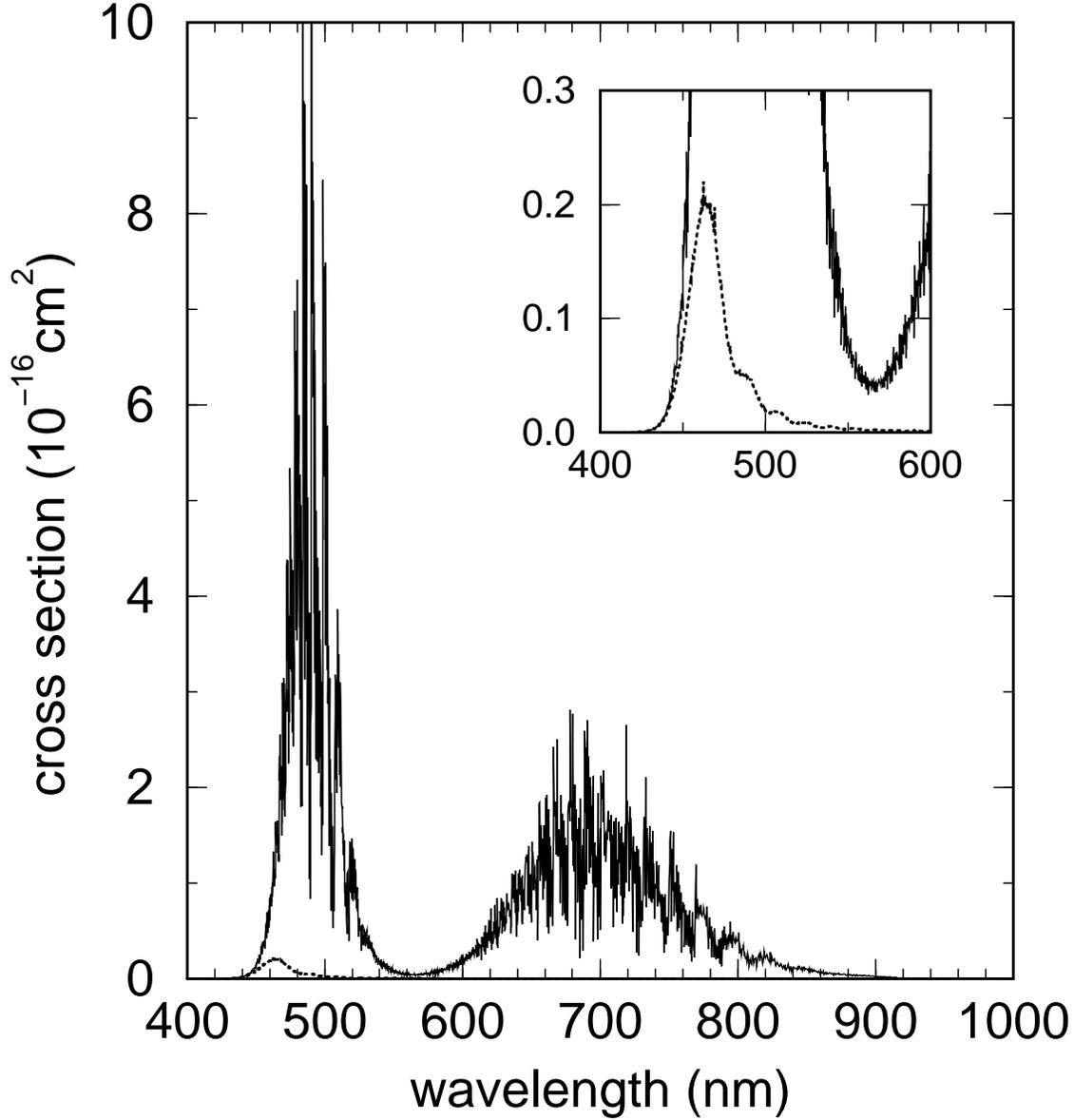}
\caption{Total absorption cross sections from
$\mbox{X}\,{}^1\Sigma_g^+$--$\mbox{A}\,{}^1\Sigma_u^+$ and
$\mbox{X}\,{}^1\Sigma_g^+$--$\mbox{B}\,{}^1\Pi_u$ transitions including
bound to bound and bound to free transitions at a temperature of
1000~K.  The satellite feature near 900~nm does not appear at this
temperature and bound-free absorption (dotted curve) is
insignificant. The inset presents a magnified view of the bound-free
contribution.
\label{s1000K}}
\end{figure}

\begin{figure}
\centering
\epsfxsize=.9\textwidth\epsfbox{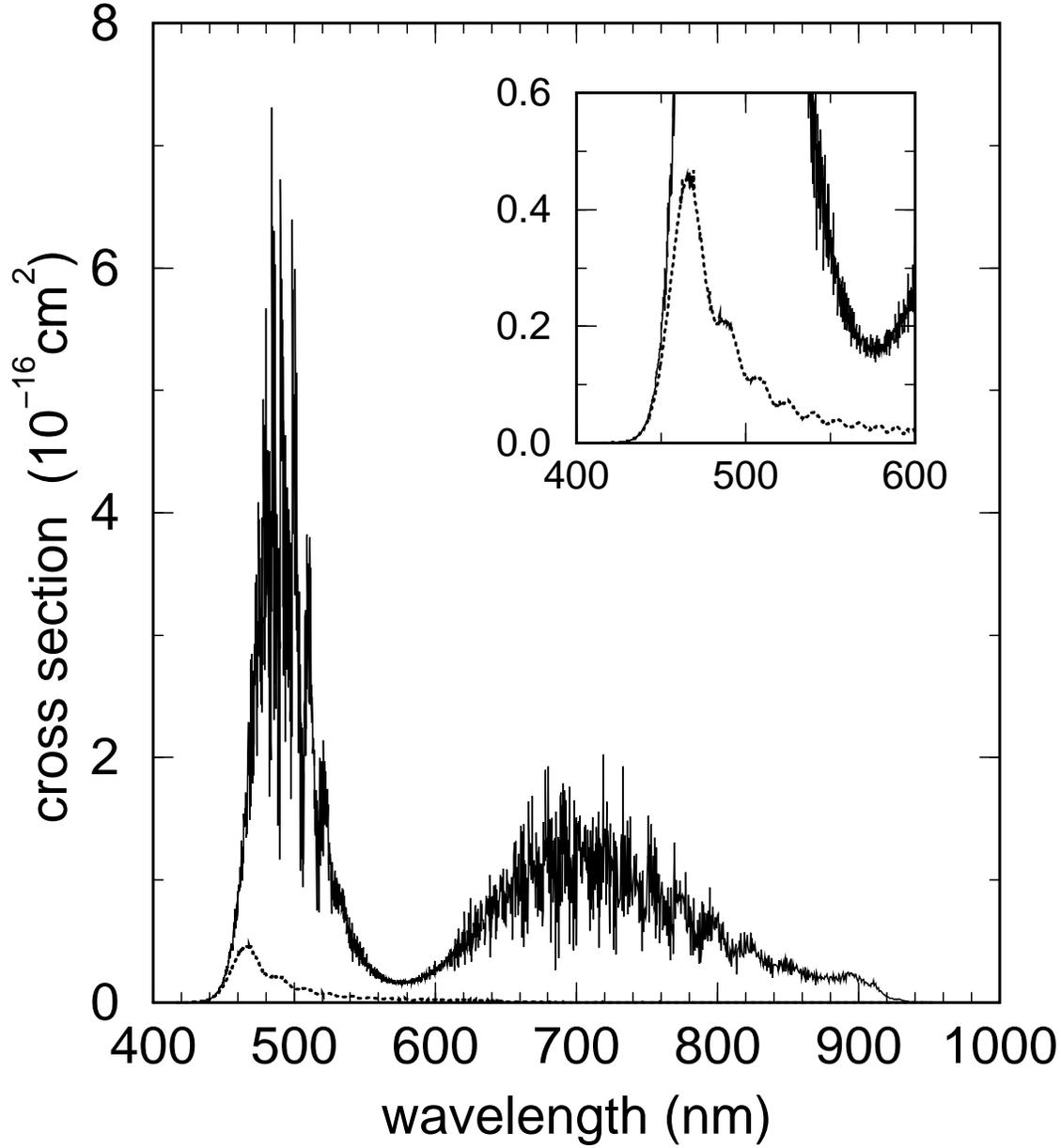}
\caption{ Total absorption cross sections from
$\mbox{X}\,{}^1\Sigma_g^+$--$\mbox{A}\,{}^1\Sigma_u^+$ and
$\mbox{X}\,{}^1\Sigma_g^+$--$\mbox{B}\,{}^1\Pi_u$ transitions including
bound to bound and bound to free transitions at a temperature of
1500~K.  As the temperature increases, the absorption spectra are
distributed over a wider spectral range and the ratio of the peak
cross sections between
$\mbox{X}\,{}^1\Sigma_g^+$--$\mbox{B}\,{}^1\Pi_u$ and
$\mbox{X}\,{}^1\Sigma_g^+$--$\mbox{A}\,{}^1\Sigma_u^+$ bands
decreases.  The satellite feature near 900~nm and bound-free
absorption (dotted curve) are noticeable at this temperature. The
inset presents a magnified view of the bound-free contribution.
\label{s1500K}}
\end{figure}

\begin{figure}
\centering
\epsfxsize=.9\textwidth\epsfbox{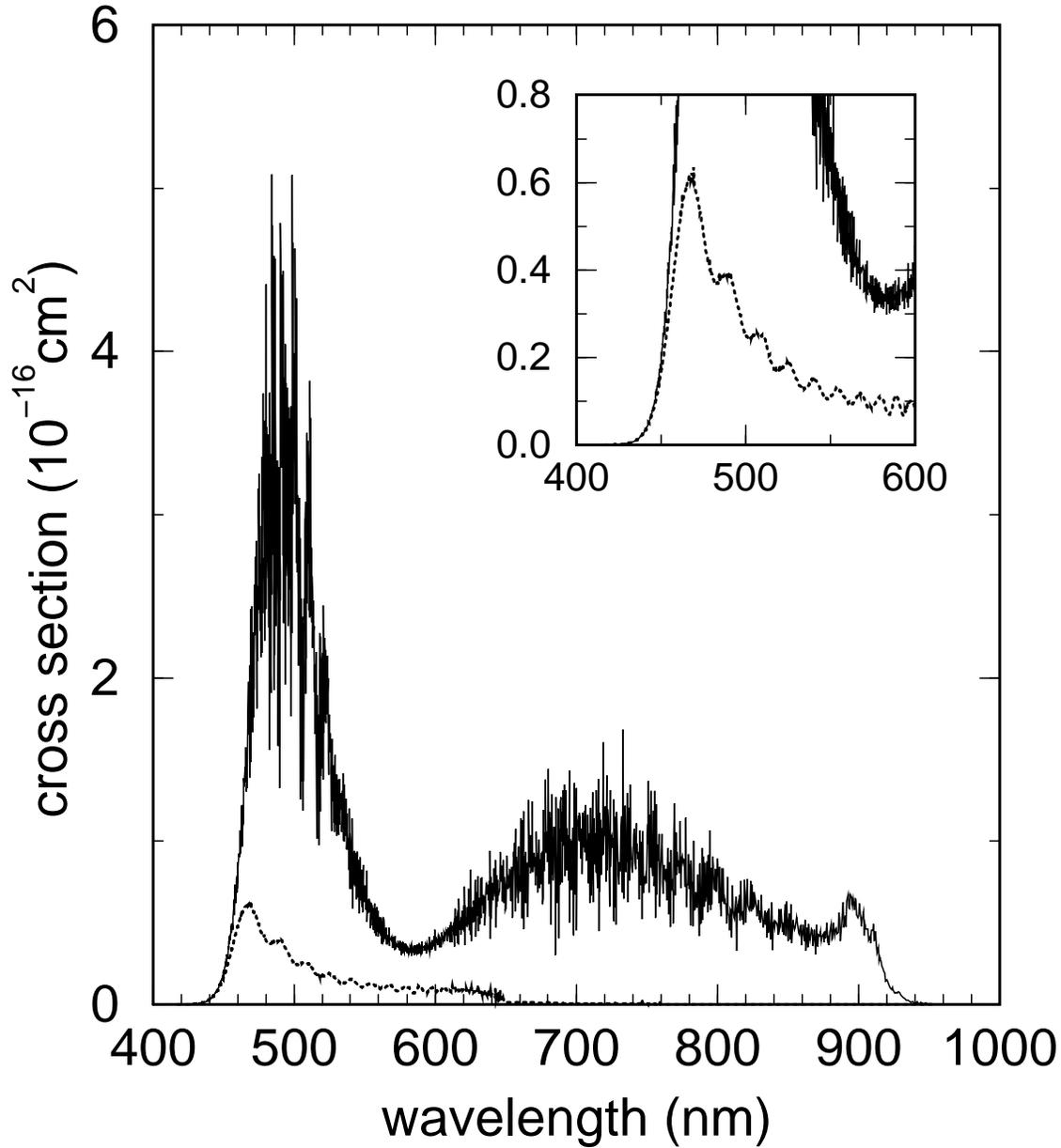}
\caption{ Total absorption cross sections from
$\mbox{X}\,{}^1\Sigma_g^+$--$\mbox{A}\,{}^1\Sigma_u^+$ and
$\mbox{X}\,{}^1\Sigma_g^+$--$\mbox{B}\,{}^1\Pi_u$ transitions
including bound to bound and bound to free transitions at a
temperature of 2033~K.  At this temperature the satellite feature near
900~nm is now prominent and bound-free absorption (dotted curve)
contributes significantly at 450~nm. The inset presents a magnified
view of the bound-free contribution.
\label{s2033K}}
\end{figure}

Absorption spectra arising from molecular singlet transitions at the
far wings of the atomic 2p line at 671~nm consist of a blue wing due
to $\mbox{X}\,{}^1\Sigma_g^+$-$\mbox{B}\,{}^1\Pi_u$ transitions and a
red wing due to $\mbox{X}\,{}^1\Sigma_g^+$-$\mbox{A}\,{}^1\Sigma_u^+$
transitions.  Calculations for bound--bound (bb) and bound--free (bf)
absorption cross sections were carried out separately using
Eq.~(\ref{e:lam}) with $\Delta\nu$ = 8 $\mbox{cm}^{-1}$.  The results
for the total (the sum of bb and bf) absorption cross sections at
temperatures of 1000~K, 1500~K and 2033~K are given in
Figs.~\ref{s1000K}--\ref{s2033K}.  The ratios of the peak cross
sections of the $\mbox{X}\,{}^1\Sigma_g^+$--$\mbox{B}\,{}^1\Pi_u$ wing
to those of the $\mbox{X}\,{}^1\Sigma_g^+$--$\mbox{A}\,{}^1\Sigma_u^+$ wing
are higher at lower temperatures.  As temperature increases,
absorption spectra spread out from the peak spectral regions and there
emerges near 900~nm a satellite feature arising from the
minimum~\cite{Jab45} in the
$\mbox{X}\,{}^1\Sigma_g^+$-$\mbox{A}\,{}^1\Sigma_u^+$ difference
potential and the maximum of the transition dipole moment
function~\cite{Woe81,RatFisKon87}.  We also show the bf contributions
to the cross sections on each plot. The bf component contributes
mainly to the extreme blue part of the blue wing and increases in
magnitude significantly as the temperature increases.  It is found
that transitions to quasibound and continuum levels of the
$\mbox{B}\,{}^1\Pi_u$ state contribute significantly to the total
absorption spectra in the case of
$\mbox{X}\,{}^1\Sigma_g^+$-$\mbox{B}\,{}^1\Pi_u$ transitions,
apparently because there is less vibrational oscillator strength
density in the discrete part of the spectrum for the B state compared
to the A state.  Transitions into quasibound states of the
$\mbox{A}\,{}^1\Sigma_u^+$ or $\mbox{B}\,{}^1\Pi_u$ states have been
included in the results for the total cross sections in
Figs.~\ref{s1000K}--\ref{s2033K}.

Theoretical quantum-mechanical calculations for absorption cross
sections from the $\mbox{X}\,{}^1\Sigma_g^+$ state to the
$\mbox{A}\,{}^1\Sigma_u^+$ state over the spectral range 600--950~nm
were carried out by Lam {\em et al.\/}~\cite{LamGalHes77} using a
constant dipole moment of 6.58~D at 1020~K, for which bf transitions
are not significant.  Our calculations in Fig.~\ref{s1000K}
are about a factor of 10 less
than the result shown in Fig.~5 of Ref.~\cite{LamGalHes77}, but agree
well both in overall shape and in details of finer structures.  We
repeated the calculations using the constant dipole moment of
Ref.~\cite{LamGalHes77} for both classical and quantum-mechanical
cross sections and although these two results agreed with each other,
they were also a factor of 10 less than the result shown in Fig.~5 of
Ref.~\cite{LamGalHes77}.  
Thus it appears to us that there may be a
mislabeling of the vertical axis in Fig.~5 of Lam {\em et al.\/} [A
similar calculation that we performed~\cite{ChuKirBab-un} for 
$\mbox{Na}_2$ at 800~K is in complete agreement with Fig.~4 of
Ref.~\cite{LamGalHes77}.]

Calculations of absorption spectra at 2033~K over the spectral range
450--750~nm presented in Fig.~\ref{s2033K} are in good agreement with
the measured values of Erdman {\em et al.\/}~\cite{ErdSanStw96} and
with quantum-mechanical calculations performed by
Mills~\cite{Mil94}. The experimental study of Erdman {\em et
al.\/}~\cite{ErdSanStw96} involved an investigation of molecular
triplet states~\cite{ErdSanStw96} and did not explore the satellite
feature at 900~nm.  The calculations over the range 450--750~nm by
Mills~\cite{Mil94} included triplet molecular transitions and are not
directly comparable with the present results.  Nevertheless since the
singlet transitions dominate the absorption we find excellent
qualitative agreement with the calculations presented by Mills.

\acknowledgements 
We thank R.~C\^ot\'e for generously sharing assembled data and
A.~Dalgarno for helpful discussions.  We also are grateful to
A. Gallagher, W. Stwalley, and M. Fajardo for helpful
correspondence. This work is supported in part by the National Science
Foundation under grant PHY97-24713 and by a grant to the Institute for
Theoretical Atomic and Molecular Physics at Harvard College
Observatory and the Smithsonian Astrophysical Observatory.

\begin{table}
\caption{Comparison of
calculated lifetimes in ns for
ro-vibrational levels of the $\mbox{A}\,{}^1\Sigma_u^+$ state. 
\label{t:Alife}}
\begin{tabular}{cccccc}
 $v'$& $J'$ &   Watson~\protect\cite{Wat77} &
 Sangfeldt {\em et al.\/}~\protect\cite{SanKurEla84} & This~work \\
\hline
  7 &  15 & 16.8  &  19.25   &   18.55 \\
  9 &  5  & 16.9  &  19.29   &   18.65 \\
 20 &  8  & 17.3  &	     &   19.04 
\end{tabular}
\end{table}
%
\begin{table}
\caption{
Lifetimes in ns for ro-vibrational levels 
of the $\mbox{A}\,{}^1\Sigma_u^+$ state calculated as described in the text.
\label{t:Alife2}}
\begin{tabular}{cccc}
$v'$ & $J'= 0$ &   $J'=9$   & $J'=15$ \\
\hline 
  0  & 17.74   &    17.77   &   17.82 \\
  1  & 17.87   &    17.90   &   17.94 \\
  2  & 17.98   &    18.01   &   18.06 \\
  3  & 18.09   &    18.12   &   18.17 \\
  4  & 18.20   &    18.23   &   18.27 \\
  5  & 18.30   &    18.33   &   18.37 \\
  6  & 18.39   &    18.42   &   18.46 \\
  7  & 18.48   &    18.51   &   18.55 
\end{tabular}
\end{table}

\clearpage
\begin{table}
\caption{Comparison of
calculated lifetimes in ns for
ro-vibrational levels of the $\mbox{B}\,{}^1\Pi_u$ state. 
\label{t:tlifetimes}}
\begin{tabular}{cccccc}
 $v'$& $J'$ &   Uzer {\em et al.\/}~\protect\cite{UzeWatDal78} &
 Sangfeldt {\em et al.\/}~\protect\cite{SanKurEla84} & This~work \\
\hline
  0 & 15 & 8.3  & 6.83 & 7.66 \\
  5 &  9 & 8.5  & 7.20 & 7.95 \\
\end{tabular}
\end{table}
\begin{table}
\caption{
Lifetimes in ns for vibrational-rotational levels 
of the $\mbox{B}\,{}^1\Pi_u$ state calculated as described in the text.
\label{t:tlifetimes2}}
\begin{tabular}{cccc}
$v'$ &  $J'= 1$ & $J'=9$ & $J'=15$ \\
\hline 
0    &    7.65  &   7.65 & 7.66 \\
1    &    7.70  &   7.71 & 7.72 \\
2    &    7.76  &   7.76 & 7.77 \\
3    &    7.81  &   7.82 & 7.83 \\
4    &    7.88  &   7.89 & 7.90 \\
5    &    7.94  &   7.95 & 7.97 \\
6    &    8.02  &   8.03 & 8.04 \\
7    &    8.10  &   8.10 & 8.12 
\end{tabular}
\end{table}
%
\clearpage
\clearpage
%
\clearpage
\clearpage
%
\clearpage
%
\clearpage

\end{document}